# Bulk Superconductivity in Bismuth oxy-sulfide $Bi_4O_4S_3$

Shiva Kumar Singh[#], Anuj Kumar[#], Bhasker Gahtori[#], Shruti Kirtan[$], Gyaneshwar Sharma[$], Satyabrata Patnaik[$] and Veer P.S. Awana[#,*]

[#]*Quantum Phenomena and Applications Division, National Physical Laboratory (CSIR)*
*Dr. K. S. Krishnan Road, New Delhi-110012, India*
[$]*School of Physical Sciences, Jawaharlal Nehru University, New Delhi-110067, India*

**ABSTRACT:** Very recent report[1] on observation of superconductivity in $Bi_4O_4S_3$ could potentially reignite the search for superconductivity in a broad range of layered sulfides. We report here synthesis of $Bi_4O_4S_3$ at $500^0C$ by vacuum encapsulation technique and its basic characterizations. $Bi_4O_4S_3$ is contaminated by small amounts of $Bi_2S_3$ and Bi impurities. The majority phase is tetragonal *I4/mmm* space group with lattice parameters $a$ = 3.9697(2)Å, $c$ = 41.3520(1)Å. Both *AC* and *DC* magnetization measurements confirmed that $Bi_4O_4S_3$ is a bulk superconductor with superconducting transition temperature ($T_c$) of 4.4K. Isothermal magnetization (*MH*) measurements indicated closed loops with clear signatures of flux pinning and irreversible behavior. The lower critical field ($H_{c1}$) at 2K, of the new superconductor is found to be ~15 Oe. The magneto-transport $R(T, H)$ measurements showed a resistive broadening and decrease in $T_c$ ($\rho$=0) to lower temperatures with increasing magnetic field. The extrapolated upper critical field $H_{c2}(0)$ is ~ 31kOe with a corresponding Ginzburg-Landau coherence length of ~100Å . In the normal state the $\rho \sim T^2$ is not indicated. Hall resistivity data show non-linear magnetic field dependence. Our magnetization and electrical transport measurements substantiate the appearance of bulk superconductivity in as synthesized $Bi_4O_4S_3$. On the other hand same temperature heat treated Bi is not superconducting, thus excluding possibility of impurity driven superconductivity in the newly discovered $Bi_4O_4S_3$ superconductor.

The discovery of superconductivity at 26K in $LaO_{1-x}F_xFeAs$[2] has ignited a gold rush in the search of new superconductors. Besides popular Fe based pnictides[2,3] and chalcogenides[4], some new interesting systems have also appeared. To name some, they are $CeNi_{0.8}Bi_2$[5], $BiOCuS$[6] and doped $LaCo_2B_2$[7]. The superconducting transition temperatures of these systems are around 4K. These compounds are layered with relatively large unit cells and mimic the superconducting characteristics of $CuO_2$ based HTSc cuprates and FeAs based pnictides. A comprehensive theoretical understanding of the mechanisms of $CuO_2$ and FeAs based high temperature superconductivity is still awaited. The hybridization of Cu-O and Fe-As in these strongly correlated systems along with their multiband character has been of prime interest to the scientific community[8,9]. After the recent observations of superconductivity in $BiOCuS$[6] and doped $LaCo_2B_2$[7], it is a pertinent question to ask if CuS and CoB could also play the same role as of $CuO_2$ and FeAs. In this regards, it is worth mentioning that although superconductivity of BiOCuS could not be reproduced[10], the $CeNi_{0.8}Bi_2$ and doped $LaCo_2B_2$ still lack independent confirmation. For example the volume fraction of superconductivity in $CeNi_{0.8}Bi_2$ is very small[11]. In this sense, the observation of superconductivity at around 4K in $Bi_4O_4S_3$[1] has once again started the debate; whether this newest series of superconductivity is intrinsic or not. It is suggested that superconductivity of $Bi_4O_4S_3$ is $BiS_2$ based and doping mechanism is similar to that of cuprates and pnictides[12,13]. The central question is whether the observed superconductivity in $Bi_4O_4S_3$ is intrinsic or it is being triggered by Bi impurity in the matrix.

Bismuth has been a part of various superconducting compounds, such as Bi based High Temperature cuprates (BSSCO)[14], $Bi_3Ni$[15,16] and $CeNi_{0.8}Bi_2$[5] compounds. On the other hand, pure Bismuth is found in several phases, out of which ordinary rhombohedral Bi phase is non-superconducting[17,18], while some other phases are found to be superconducting[19-23]. Various crystallographic phases of pure Bi, which are superconducting in the bulk phase, are Bi II, III and V (high-pressure phases of Bi) with $T_c$ = 3.9K, 7.2K, and 8.5K[19-21] respectively. The *fcc* Bi phase superconducts with $T_c \sim$ 4K[22]; and amorphous Bi with $T_c$ = 6K[23].

In the current communication, we report on the extensive characterization of the newly discovered[1] $Bi_4O_4S_3$ superconductor. The synthesized $Bi_4O_4S_3$ is crystallized in tetragonal structure with space group *I4/mmm*. The main phase of the sample is contaminated with small impurities of Bi and $Bi_2S_3$. $Bi_4O_4S_3$ compound is found to be bulk superconducting at around 4.4K, as confirmed from magnetization and transport measurements. Interestingly same route heat treated pure Bi is non-superconducting. Bi is in rhombohedral phase and hence is non-superconducting. Our results conclude that superconductivity of $Bi_4O_4S_3$ is intrinsic and not driven by Bi impurity phase.

$Bi_4O_4S_3$ was synthesized by solid state reaction route via vacuum encapsulation. High purity Bi, $Bi_2O_3$ and S were weighed in right stoichiometric ratio and ground thoroughly in the glove box under high purity argon atmosphere. The powders were subsequently pelletized and vacuum-sealed ($10^{-4}$ Torr) in separate quartz tubes. Sealed quartz ampoules were placed in box furnace and heat treated at $500^0C$ for 18h followed by cooling to room temperature naturally. The process was repeated twice. The *X*-ray diffraction (*XRD*) pattern of the compounds was recorded on Rigaku diffractometer. Rietveld refinement of *XRD* pattern was carried out using *FullProf* software. The magnetization and transport measurements were carried out using 14 Tesla *Cryogenic PPMS* (Physical Property Measurement System).

The synthesized $Bi_4O_4S_3$ sample is gray in color. On the other hand, Bi sample is of shiny silver color. The room temperature *XRD* pattern for synthesized Bi and $Bi_4O_4S_3$ samples are shown in Figure l (a).

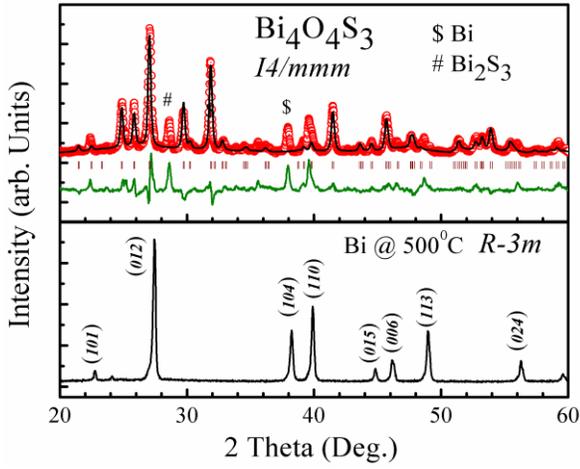

*Figure 1 (a).* Rietveld refined room temperature *X*-ray diffraction (*XR*D) patterns of $Bi_4O_4S_3$ and same temperature heat treated Bi.

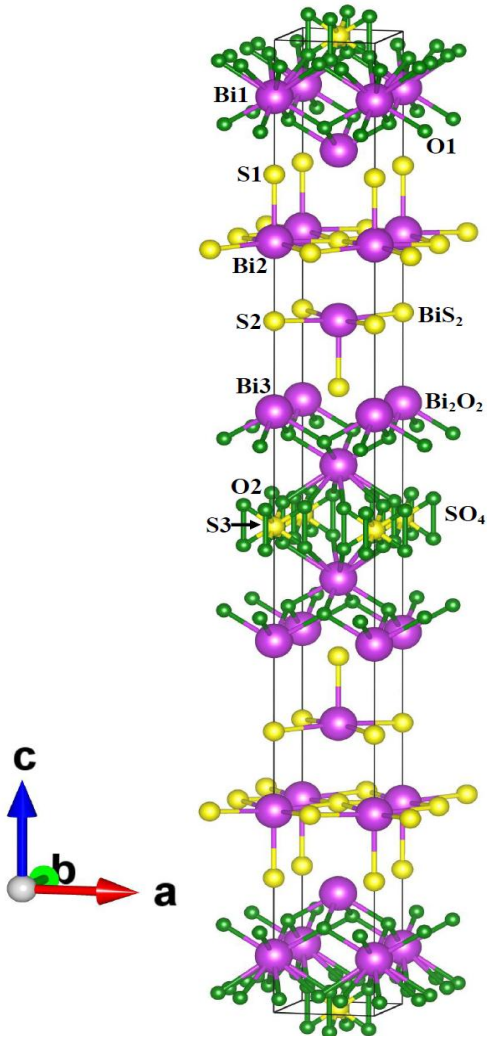

*Figure 1 (b).* The schematic unit cell of the $Bi_4O_4S_3$ compound. Color code: Voilet-Bismuth, Yellow-Sulfur and Green-Oxygen.

The $Bi_4O_4S_3$ sample is crystallized in tetragonal structure with space group *I4/mmm*. Rietveld refinement of *XR*D patterns are carried out using reported[1] Wyckoff positions. The positions and lattice parameters are further refined. The lattice parameters thus obtained are $a = 3.9697 (2)$Å and $c = 41.3520 (1)$Å. The Wyckoff positions of the $Bi_4O_4S_3$ compound are given in Table 1. The *XR*D pattern of same temperature heat treated Bi, is also depicted in Figure 1 (a), which is crystallized in clean rhombohedral phase. It can be concluded from *XR*D results that the synthesized $Bi_4O_4S_3$ sample is nearly single phase with some impurities of rhombohedral Bi and $Bi_2S_3$. Rhombohederal Bi is reported to be non-superconducting[17,18].

The structure of $Bi_4O_4S_3$ is still under debate within the space group crystallization of *I4/mmm* or *I -42m*[1]. The representative unit cell of the compound in *I4/mmm* space group crystallization is shown in Figure 1 (b). The layered structure includes $Bi_2S_4$ (rock-salt type), $Bi_2O_2$ (fluorite type) and $SO_4$ layers. Superconductivity is induced in $BiS_2$ layer due to Bi-6*p* and S-3*p* orbitals hybridization. The theoretical calculations[13] show that, bands are derived from Bi-6*p* and in-plane S-3*p* orbitals. These are dominating bands for electron conduction and superconductivity.

| *Atom* | *x* | *y* | *z* | *site* | *Fractional occupancy* |
|---|---|---|---|---|---|
| *Bi1* | 0.0000 | 0.0000 | 0.0583 (4) | *4e* | 1 |
| *Bi2* | 0.0000 | 0.0000 | 0.2074 (2) | *4e* | 1 |
| *Bi3* | 0.0000 | 0.0000 | 0.3821 (2) | *4e* | 1 |
| *S1* | 0.0000 | 0.0000 | 0.1383 (1) | *4e* | 1 |
| *S2* | 0.0000 | 0.0000 | 0.2890(1) | *4e* | 1 |
| *S3* | 0.0000 | 0.0000 | 0.5000 | *2b* | 1/2 |
| *O1* | 0.0000 | 0.5000 | 0.0884(1) | *8g* | 1 |
| *O2* | 0.0000 | 0.3053(1) | 0.4793(2) | *16n* | 1/4 |

*Table 1.* Rietveld refined Wyckoff positions and fractional occupancies of the atoms in $Bi_4O_4S_3$.

Various atoms with their respective positions are labeled in Figure 1(b) and their coordinates are provided in Table 1. Bismuth (Bi1, Bi2 and Bi3) and Sulfur (S1 and S2) atoms occupy the *4e* (0, 0, *z*) site. S3 atom is at *2b* (0, 0, ½) site. O1 is situated at *8g* (0, ½, *z*) and the O2 atom is positioned at *16n* (0, *y*, *z*) site. The structural refinement indicates that the molecular composition is $Bi_3O_3S_{2.25}$. It is the normalization of $Bi_4O_4S_3$ composition by ¾. The superconducting phase (*i.e.* $Bi_4O_4S_3$ or $Bi_3O_3S_{2.25}$) is 25% $SO_4$ deficient composition of the parent $Bi_3O_4S_{2.5}$ ($Bi_6O_8S_5$) compound[1].

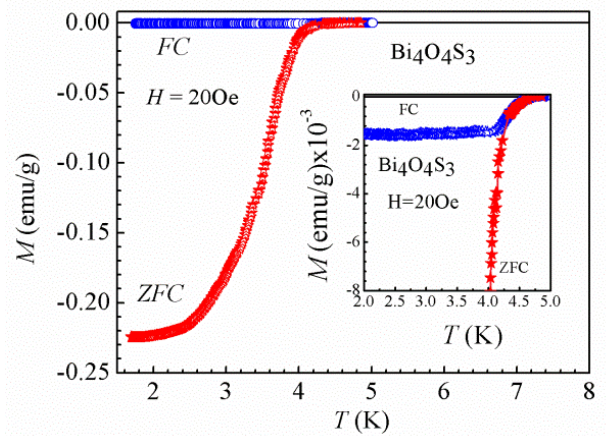

*Figure 2.* Temperature variation of D*C* Magnetization in *ZFC* and *FC* mode for $Bi_4O_4S_3$ compound at 20Oe, onset $T_c$ is identified at 4.4K. Inset shows the expanded part of the same plot indicating irreversible behavior.

The *DC* magnetic susceptibility of $Bi_4O_4S_3$ sample is shown in Figure 2. The magnetization is done in both, *FC* (Field cooled) and *ZFC* (Zero-field-cooled) protocols under applied magnetic field of 20Oe. The compound shows sharp superconducting onset at 4.4K, which is clear from the zoomed inset of Figure 2. There is evidence for substantial flux trapping too. The bifurcation of *FC* and *ZFC* below $T_c$ marks the irreversible region. The shielding fraction as evidenced from *ZFC* diamagnetic susceptibility is ~95%. Both *FC* and *ZFC* magnetization data confirm the appearance of bulk superconductivity in $Bi_4O_4S_3$. In order to exclude the role of Bi impurity in superconductivity of $Bi_4O_4S_3$, we measured the magnetization of the same temperature ($500^0C$) heat treated Bi and found the same to be non-superconducting (plot not shown). It can be seen in Figure 1, that the $500^0C$ heat treated Bi is crystallized in non-superconducting[17,18] rhombohedral phase. This excludes the possibility of un-reacted Bi driven superconductivity in $Bi_4O_4S_3$. In fact the sufficient superconducting volume fraction and large shielding (~95%) of our studied sample, discards the possibility of the minor impurity phase driven superconductivity.

The *AC* susceptibility versus temperature $\chi(T)$ behavior of the $Bi_4O_4S_3$ sample is exhibited in Figure 3. *AC* susceptibility is done at 1kHz and 10Oe *AC* drive field. *DC* applied field is kept zero to check the superconducting transition temperature and is increased to 5kOe and 10kOe to further check the *AC* loses in the mixed state. Both the real ($\chi'$) and imaginary ($\chi''$) part of AC susceptibility were measured. Real part ($\chi'$) susceptibility shows sharp transition to diamagnetism at around 4.4K, confirming bulk superconductivity. The imaginary part on the other hand exhibits a single sharp peak in positive susceptibility at around the same temperature. Presence of single sharp peak in $\chi''$ is reminiscent of better superconducting grains coupling in studied $Bi_4O_4S_3$ superconductor. Under applied *DC* field of 5kOe the $\chi'$ diamagnetic transition is shifted to lower temperature of 2.6K and the corresponding $\chi''$ peak is broadened and shifted to same lower temperature. This is usual for a type-II superconductor. At 10kOe *DC* field neither $\chi'$ nor $\chi''$ show any transitions, indicative of rapid suppression of superconductivity.

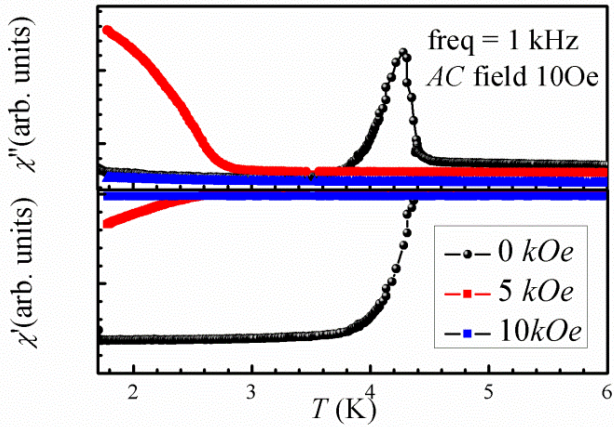

*Figure 3.* AC susceptibility $\chi(T)$ behavior of the $Bi_4O_4S_3$ sample at frequency 1 kHz and *AC* drive amplitude 10Oe under various (0, 5, 10kOe) *DC* applied fields.

Figure 4 shows the isothermal *MH* curve of the sample at 2K, up to applied field of 3kOe. Upper inset of the figure shows the same up to 1kOe. The *MH* curve (lower inset) shows that the initial flux penetration and the deviation from linearity marks lower critical field ($H_{c1}$) of this compound ~15 Oe (at 2K). Wide open *MH* loop of the studied $Bi_4O_4S_3$ compound demonstrates bulk superconductivity.

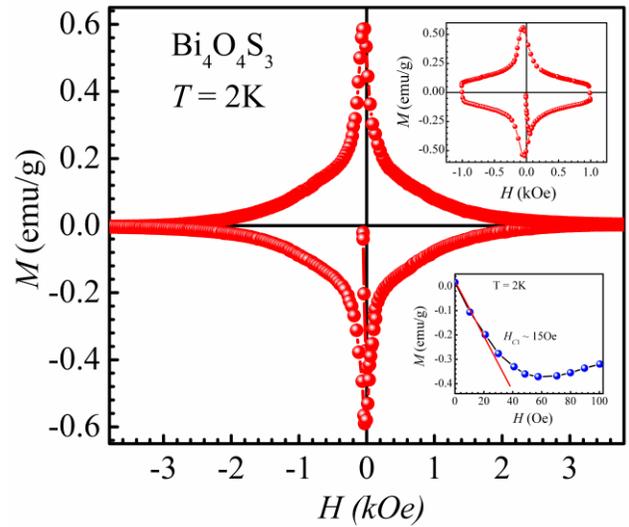

*Figure 4.* Isothermal magnetization with field (*MH*) at 2K in an applied field up to 3kOe. Insets of the figure show the same in smaller field ranges. $H_{c1}$ (2K) is estimated to be 15Oe.

Figure 5 depicts the resistivity versus temperature ($\rho$-*T*) measurement with and without applied magnetic field. The resistivity of the sample decreases with temperature and confirms superconductivity with onset $T_c$ ~4.4 K. The normal state conduction (inset Figure 5) is of metallic type and a $T^2$ fitting is found to be inappropriate implying non-Fermi liquid behavior. With applied field of 1, 2 and 5kOe, the $T_c$ ($\rho = 0$) decreases to 3.2, 2.7 and 2K. With further higher fields of 10 and 20kOe, the $T_c$ ($\rho = 0$) state is not observed and only $T_c$ (onset) is seen. As sketched in Figure 5, we have estimated upper critical field $H_{c2}$ (*T*) by using the conservative procedure of intersection point between linear slope lines of normal state resistivity and superconducting transition line. While the applicability of *WHH* (Werthamer-Helfand-Hohenberg) approximation can be debated in this new superconductor, a simplistic single band extrapolation leads to $H_{c2}(0)$ (= - 0.69 $T_c$ $dH_{c2}/dT|_{Tc}$) value of 31kOe. From this the Ginzburg-Landau coherence length $\xi = (2.07 \times 10^{-7}/2\pi H_{c2})^{1/2}$ is estimated to be ~100 Å.

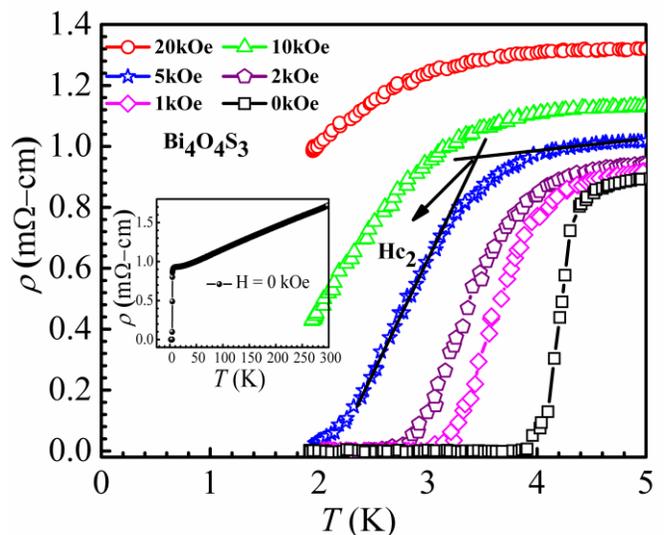

*Figure 5.* Resistivity vs. temperature ($\rho$-*T*) behavior of $Bi_4O_4S_3$ in various applied fields of 0, 1, 2, 5, 10 and 20kOe in superconducting region; inset shows the zero field $\rho$-*T* in extended temperature range of 2-300K.

A strong magneto-resistance in the normal state is also seen that can possibly be ascribed to reasons similar to extra Mg impurity in $MgB_2$ i.e. due to Bi impurity in the matrix. It is predicted by the theoretical calculations that superconductivity in $BiS_2$ layers is of multiband type[13,24]. In Figure 6 we plot the Hall resistivity as a function of magnetic field at 10K. The dominance of electronic charge carrier in normal state conduction mechanism is confirmed. Strong non-linearity is observed with increasing magnetic field that is suggestive of the deviations from single band analysis.

The Hall coefficient [Figure 6 Inset (a)] is field dependent and the carrier concentration at low field is estimated to be ~$1.53 \times 10^{19}$ per $cm^3$ at 10K that increases to ~$2.4 \times 10^{19}$ per $cm^3$ at 300K.

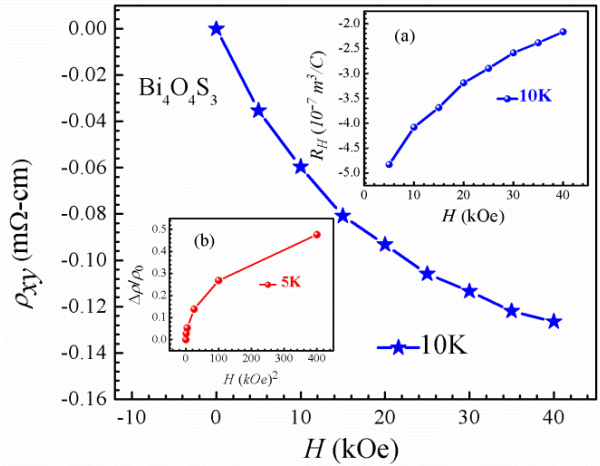

*Figure 6.* Hall resistivity is plotted as a function of magnetic field at $T = 10K$. Inset (a) shows variation of Hall coefficient as a function of field. Inset (b) shows normalized magneto-resistance at 5K, implying non-$H^2$ dependence.

In the inset (b) of Figure 6, we show $\Delta\rho(H)/\rho(0)$ versus $H^2$ at 5K for fields up to 20kOe. One of the established features of multiband superconductivity is the dependence $\Delta\rho(H) \propto H^2$ in low field range. Evidently, in this regime, this dependence is not seen. Taken in totality, we can conclude that while our Hall resistivity data may demand incorporation of more rigorous analysis, the magneto-resistance aspects in $Bi_4O_4S_3$ could be well due to Bi impurity.

In conclusion we have synthesized the new layered sulfide $Bi_4O_4S_3$ superconductor and established its bulk superconductivity by magnetization and transport measurements. Detailed Reitveld analysis determines the molecular composition as $Bi_3O_3S_{2.25}$. The coherence length is estimated to be ~100Å. A departure from strong electron-electron correlation in the normal state is indicated. The Hall resistivity yields non-linear magnetic field dependence.

## AUTHOR INFORMATION

*Corresponding Author

Dr. V. P. S. Awana, Senior Scientist
E-mail: awana@mail.npindia.org
Ph. +91-11-45609357, Fax-+91-11-45609310

### Author Contributions

All authors contributed equally.

## ACKNOWLEDGMENT

This work is supported by DAE-SRC outstanding investigator award scheme on search for new superconductors. Authors from NPL would like to thank their Director Prof. R.C. Budhani for his keen interest in the present work. Shiva Kumar and Anuj Kumar are thankful to CSIR-India for providing the financial support during their research. Shruti and G. Sharma acknowledge UGC for research fellowships. S. Patnaik thanks AIRF, JNU for the PPMS facility.

## REFERENCES

1. Mizuguchi, Y.; Fujihisa, H; Gotoh, Y.; Suzuki, K.; Usui, H.; Kuroki, K.; Demura, S.; Takano, Y.; Izawa, H.; Miura, O. arXiv:1207.3145.
2. Kamihara, Y.; Watanabe, T.; Hirano M.; Hosono, H. J. Am. Chem. Soc. **2008** *130* 3296.
3. Ren, Z. A.; Lu, W.; Yang, J.; Yi, W.; Shen, X. L.; Li, Z. C.; Che, G. C.; Dong, X. L.; Sun, L. L.; Zhou F.; Zhao, Z. X. Chin. Phys. Lett. **2008** *25* 2215.
4. Hsu, F. C.; Luo, J. Y.; Yeh, K. W.; Chen, T. K.; Huang, T. W.; Wu, P. M.; Lee, Y. C.; Huang, Y. L.; Chu, Y. Y.; Yan D. C.; Wu, M. K. Proc. Natl. Acad. Sci. U.S.A **2008** *105* 14262.
5. Mizuguchi, H.; Matsuishi, S.; Hirano, M.; Tachibana, M.; -Muromachi, E. T.; Kawaji H.; Hosono, H. Phys. Rev. Lett. **2011** *106* 057002.
6. Ubaldini, A.; Giannini, E.; Senatore, C.; van de Marel, D. Physica C **2010** *470* S356.
7. Mizuguchi, H.; Kuroda, T.; Kamiya, T.; Hosono, H. Phys. Rev. Lett. **2011** *106* 0277001.
8. Pickett, W. E. Rev. Mod. Phys **1989** *61* 433.
9. Mazin I. I.; Singh, D. J.; Johannes M. D.; Du, M. H. Phys. Rev. Lett. **2008** *101* 057003.
10. Pal, A.; Kishan, H.; Awana, V. P. S. J. Supercond. & Novel Magn. **2010** *23* 301.
11. Kumar, A.; Kumar, S.; Jha, R.; Awana, V. P. S. J. Supercond. & Novel Magn. **2012** *25* 723.
12. Mizuguchi, Y.; Demura, S.; Deguchi, K.; Takano, Yoshihiko.; Fujihisa, H.; Gotoh, Y.; Izawa, H.; Miura, O. arXiv:1207.3558.
13. Usui, H.; Suzuki, K.; Kuroki, K. arXiv:1207.3888.
14. Maeda, H.; Tanaka, Y.; Fukutumi, M.; Asano T. Jpn. J. Appl. Phys. **1988** *27* L209.
15. Alekseevskii, N. E.; Brandt N. B.; Kostina, T. I. J. Exp. Theor. Phys. **1951** *21* 951.
16. Kumar, J.; Kumar, A.; Vajpayee, A.; Gahtori, B.; Sharma, D.; Ahluwalia, P. K.; Auluck, S.; Awana, V. P. S. Sup. Sci. and Tech. **2011** *24* 085002.
17. Muntyanu, F. M.; Gilewski, A.; Nenkov, K.; Warchulska, J.; Zaleski, A. J. Phys. Rev. B **2006** *73* 132507.
18. Wetzel B.; Micklitz H., Phys. Rev. Lett. **1991** *66* 385.
19. Chester P. F.; Jones, G. O. Philos. Mag. **1953** *44* 1281.
20. Brandt N. B.; Ginzburg, N. J.; Zh. Eksp. Teor. Fiz. **1961** *39* 1554.
21. Wittig, J. Z. Phys. **1966** *195* 228.
22. Moodera, J. S.; Meservey, R. Phys. Rev. B **1990** *42* 179.
23. Buckel W.; Hilsch, R. Z. Phys. **1954** *138* 109.
24. Yang S. L.; Tao J.; Ding X.; Wen H. H. arXiv:1207.4955